\documentclass[usenatbib,fleqn]{mnras}
\usepackage{newtxtext,newtxmath}

\usepackage[T1]{fontenc}

\DeclareRobustCommand{\VAN}[3]{#2}
\let\VANthebibliography\thebibliography
\def\thebibliography{\DeclareRobustCommand{\VAN}[3]{##3}\VANthebibliography}

\usepackage{graphicx}	
\usepackage{amsmath}	
\usepackage{bm}

\makeatletter
\newlength{\abovecaptionskip}%
\makeatother
\usepackage{threeparttable}
\usepackage{verbatim}
\usepackage{epstopdf}
\usepackage{hyperref}
\graphicspath{{./figures/}}
\usepackage{url}

\newcommand\lsim{\mathrel{\rlap{\lower4pt\hbox{\hskip1pt$\sim$}}
    \raise1pt\hbox{$<$}}}
\newcommand\gsim{\mathrel{\rlap{\lower4pt\hbox{\hskip1pt$\sim$}}
    \raise1pt\hbox{$>$}}}

\def\agnote#1{{#1}}
\defcitealias{generozov&madigan2020}{GM20}
\graphicspath{{./}{figures/}}

\title[Mass ratio and the hills mechanism]{Mass ratio, the hills mechanism, and the galactic centre S-stars}
\author[A. Generozov]{Aleksey Generozov\thanks{alge9397@colorado.edu}$^{1}$\\$^{1}$JILA and the Astrophysical and Planetary Sciences Department\\
University of Colorado, Boulder}

\date{Accepted 2020 December 10. Received 2020 December 9; in original form 2020 October 13}

\pubyear{2020}

\begin{document}
\label{firstpage}
\pagerange{\pageref{firstpage}--\pageref{lastpage}}
\maketitle

\begin{abstract}
The Galactic centre contains several young populations within its central parsec: a disk between $\sim$0.05 and 0.5 pc from the centre, and the isotropic S-star cluster extending an order of magnitude further inwards in radius. Recent observations (i.e. spectroscopy and hypervelocity stars) suggest that some S-stars originate in the disk. In particular, the S-stars may be remnants of tidally disrupted disk binaries. However, there is an apparent inconsistency in this scenario: the disk contains massive O and Wolf--Rayet stars while the S-stars are lower mass, B stars. We explore two different explanations for this apparent discrepancy: (i) a built-in bias in binary disruptions, where the primary star remains closer in energy to the centre-of-mass orbit than the secondary and (ii) selective tidal disruption of massive stars within the S-star cluster. The first explanation is plausible. On the other hand, tidal disruptions have not strongly affected the mass distribution of the S-stars over the last several Myr.  
\end{abstract}
\begin{keywords}
black hole physics -- binaries: general -- Galaxy: centre\end{keywords}

\section{Introduction}
\label{sec:intro}
Our Galaxy contains a cluster of early-type stars $\sim 0.01$ pc from its central supermassive black hole (SMBH), known as the ``S-stars'' \citep{genzel+1997, ghez+1998, gillessen+2017}. At these radii, tidal forces from the central SMBH pose a challenge to in-situ formation models of these stars \citep{ghez+2003}. Instead these stars may have migrated from larger scales, either via a gas disk \citep{levin2007} or via the ``Hills mechanism,'' where binaries are disrupted by the SMBH. These disruptions typically leave one bound star in a highly eccentric orbit, while the other is ejected from the Galactic nucleus as a hypervelocity star \citep{hills1988,ginsburg&loeb2006,perets+2007,lockmann+2009,madigan+2009,dremova+2019}. A number of hypervelocity stars and candidates have been identified within our Galaxy \citep{warren_brown+2005, warren_brown+2014,  boubert+2018}

In fact, recent observations suggest many of the S-stars are remnants of disrupted binaries from a larger disk structure in the Galactic centre (the ``clockwise disk''; \citealt{levin&beloborodov03,paumard+2006}). For example, the spectroscopic ages of the S-stars and the disk are consistent \citep{lu+2013, habibi+2017}. Also, hypervelocity star observations suggest the disk produced a burst of binary disruptions in the recent past. In particular, the hypervelocity star S5-HVS1 was ejected from the Galactic centre 4.8 Myr ago with a velocity of $\sim$1800 km s$^{-1}$. The flight time of this star and its velocity vector imply an association with the disk \citep{koposov+2019}. Theoretically, the disk may produce binary disruptions via a secular gravitational instability shortly after its formation \citep{madigan+2009, generozov&madigan2020}. 

 However, there is an apparent inconsistency in this scenario. Namely, the disk contains O and Wolf--Rayet stars, which are more massive than the observed S-stars. (The S-star cluster contain no stars above $15 M_{\odot}$; \citealt{habibi+2017, cai+2018}.) \citet{generozov&madigan2020} (henceforward GM20) proposed two possible solutions to this conundrum. Firstly, the mismatch could be a sampling effect as there only $\sim$20 S-stars. Secondly, the mismatch could be due to a built-in bias in binary disruptions. Namely, in an unequal-mass binary, the secondary star would get a larger energy kick and could be deposited at smaller semimajor axes.  Here, we identify the regions of parameter space where these two mechanisms would be effective. In particular, we use the observation that the S-star cluster contains no O stars to put constraints on the mass ratio distribution of the disk binaries.

Alternatively, massive O/Wolf--Rayet stars could be more easily torqued to tidal disruption, since their tidal radius is larger \citep{chen&amaro-seoane2014}. We return to this point in \S~\ref{sec:dis}. Also, some of the S-stars could be produced by scatterings between
stellar-mass black holes and binaries in the disk \citep{trani+2019}. This process favors lower mass B stars over O stars, and would explain the mass discrepancy. However, it does not account for the observed hypervelocity stars.

Observational studies have found a range of mass ratio distributions for massive star binaries. Typically, the mass ratio distribution, $f(q)$ is parameterized as a power law $f(q)\propto q^{-\alpha}$, where $q$ is the ratio of the secondary mass to the primary mass. \citet{kiminki&kobulnicky2012} find  $\alpha=-0.1 \pm 0.5$ in the Cygnus OB2 association, while \citet{sana+2013} find $\alpha=1\pm 0.4$ within the Tarantula nebula of the LMC.  
Within the Galactic centre, three O/WR binaries are known. Two of these (E60 and IRS 16NE) have mass ratios of $\sim 2$, while the third (IRS 16SW) is a ``twin'' system with a mass ratio of 1 \citep{martins+2006, pfuhl+2014}. Such systems may arise from pre-main sequence mass transfer \citep{krumholz&thompson2007}. However, IRS 16SW is currently in Roche Lobe contact, which suggests that post-main sequence mass transfer may have affected the mass ratio of this system (\citealt{martins+2006}; although they argue that the component spectra indicate significant mass transfer has not taken place yet). Future observations of the disk may provide additional constraints, considering there is indirect evidence for a much larger binary population within it. For example, binaries could account for the observed radial trend in disk membership among young stars in the Galactic centre \citep{naoz+2018}.

The remainder of this paper is organized as follows. In \S~\ref{sec:mr}, we review the effects of mass ratio on binary disruptions. 
In \S~\ref{sec:mc}, we describe the approximate method we use to model disruptions. In \S~\ref{sec:results} we describe how the probability of not producing massive S-stars varies with binary parameters. In \S~\ref{sec:dis}, we discuss the effects of stellar disruptions. Finally, we summarize our conclusions in \S~\ref{sec:conc}.

\section{Mass ratio and the hills mechanism} 
\label{sec:mr}
When an unequal mass binary is disrupted, the primary star's final orbit will be closer in energy to that of the original binary than the secondary star's. This is because the primary star is closer to the binary's centre of mass, and experiences a smaller tidal force. Analogously, when a single star is disrupted the most tightly bound debris comes from its surface \citep{rees1988}.

Stars can either gain or lose energy in the disruption, depending on the binary's phase.
The final (specific) energy of star $i$ will be (see e.g. \citealt{kobayashi+2012})
\begin{equation}
    \epsilon \approx \epsilon_o \pm \frac{G M \delta_i}{r_t^2},
\end{equation}
where $\epsilon_o$ is the initial energy of the centre-of-mass orbit, M is the mass of the SMBH, $r_t$ is the tidal radius (approximately $(M/m_{\rm bin})^{1/3} a_{\rm bin}$, where $m_{\rm bin}$ and $a_{\rm bin}$ are the binary mass and semimajor axis respectively). The $\delta_i$ factor is the characteristic distance between the star and the centre of mass. This is approximately $q a_{\rm bin}/(1+q) $ for the primary and $a_{\rm bin}/(1+q)$ for the secondary, where $q$  is the ratio of the secondary mass to the primary mass.
Thus, the post-disruption (specific) energies of the primary (star 1) and the secondary (star 2), are 

\begin{align}
    &\epsilon_1= \epsilon_o \pm k \frac{q}{1+q} \left(\frac{M}{m_{\rm bin}}\right)^{1/3} \frac{G m_{\rm bin}}{a_{\rm bin}}\nonumber\\
    &\epsilon_2= \epsilon_o \mp  k \frac{1}{1+q} \left(\frac{M}{m_{\rm bin}}\right)^{1/3} \frac{G m_{\rm bin}}{a_{\rm bin}}, 
    \label{eq:ehills}
\end{align}
where $k$ is a factor of order unity that depends on the binary's eccentricity, phase, and orientation (see \citetalias{generozov&madigan2020}). In the parabolic limit, the corresponding semimajor axes are
\begin{align}
    &a_1=\pm \frac{1}{2 k} \frac{1+q}{q} \left(\frac{M}{m_{\rm bin}}\right)^{2/3} a_{\rm bin}\nonumber\\
    &a_2=\mp \frac{1}{2 k} (1+q) \left(\frac{M}{m_{\rm bin}}\right)^{2/3} a_{\rm bin}.
\end{align}
The absolute value of the primary's semimajor axis is larger by $1/q$.

\section{Monte Carlo experiments}
\label{sec:mc}
We explore how the mass and semimajor axis distribution of the bound stars depend on the properties of the progenitor binary population. In particular, we construct mock binary samples from the distributions in Table~\ref{tab:params}. For each binary, we calculate post-disruption energies of its member stars according to equation~\eqref{eq:ehills}, assuming the initial semimajor axis of the binary's centre-of-mass orbit is 0.05 pc. The scatter from the binary's phase and orientation is encapsulated by $k$. We use the suite of $\sim 500,000$ \texttt{AR--Chain} binary disruption simulations from \citetalias{generozov&madigan2020} to construct an empirical distribution for this parameter. Although the final energies of the stars depend on the binary mass, semimajor axis, and its centre-of-mass orbit, the $k$-distribution will not depend on these properties. Thus, without loss of generality, the binary's semimajor axis is $1$ au, and the stellar masses are $10 M_{\odot}$ in these simulations. The mass of the SMBH is $4\times 10^6 M_{\odot}$.\footnote{To quadrupole order the outcome will not depend on the mass ratio between the SMBH and binary.} The binaries approach the SMBH on a parabolic orbit with a pericenter between $\left(M/m_{\rm bin}\right)^{1/3} a_{\rm bin}$ and $4 \left(M/m_{\rm bin}\right)^{1/3} a_{\rm bin}$.

From the \texttt{AR--Chain} simulation data, we determine (i) the ``effective tidal radius,'' where the probability of binary disruption is 50\% (this is $\chi (M/m_{\rm bin})^{1/3} a_{\rm bin}$, where $\chi\approx 2-3$) and (ii) the distribution of $k$ near the effective tidal radius for each of the ten binary eccentricities in our simulations. For each mock encounter in this paper, we assume the binary is disrupted at the effective tidal radius of the closest eccentricity in our simulation data. We then sample the corresponding $k$ distribution.

\begin{figure*}
    \includegraphics[width=\columnwidth]{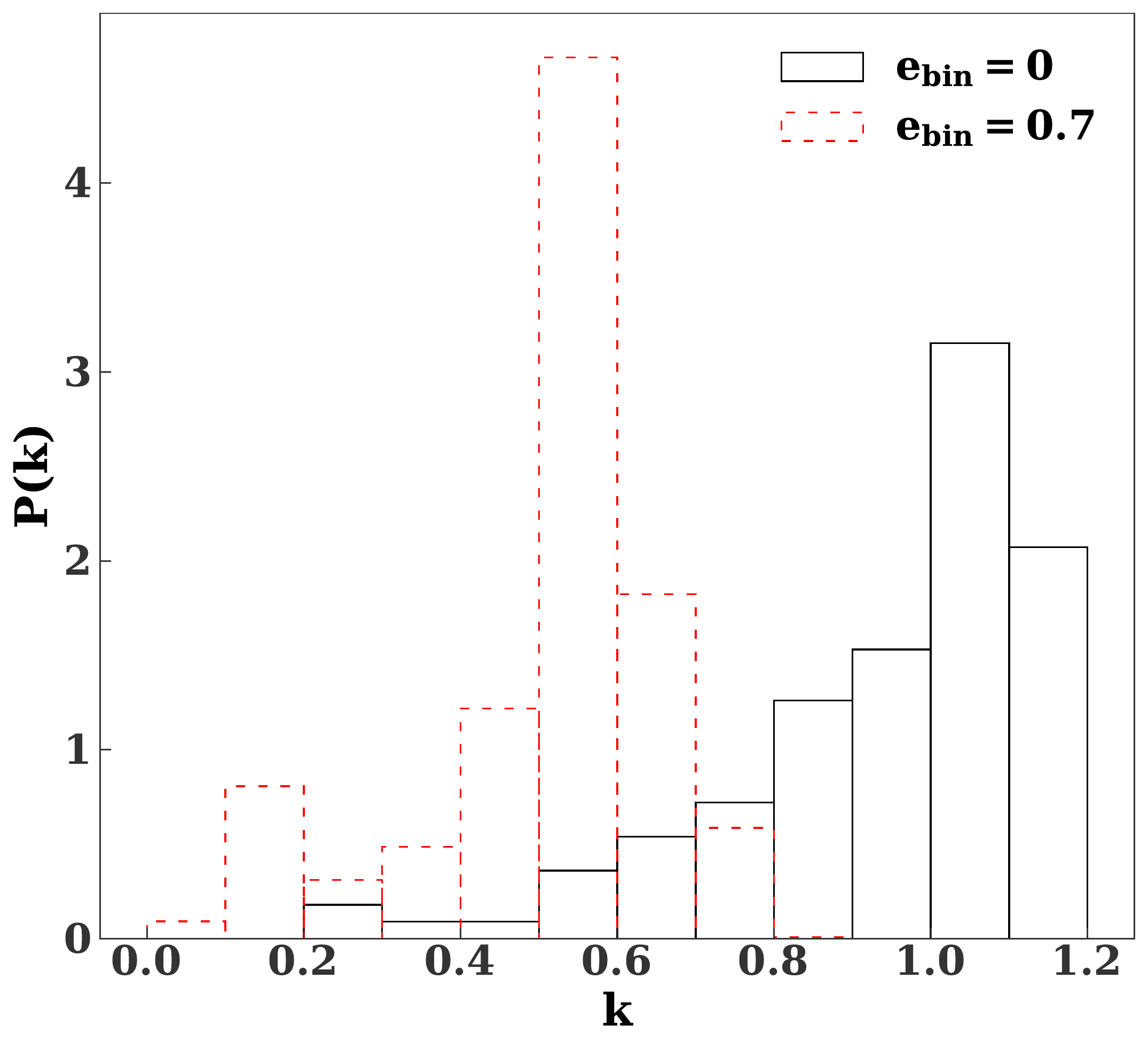}
    \includegraphics[width=\columnwidth]{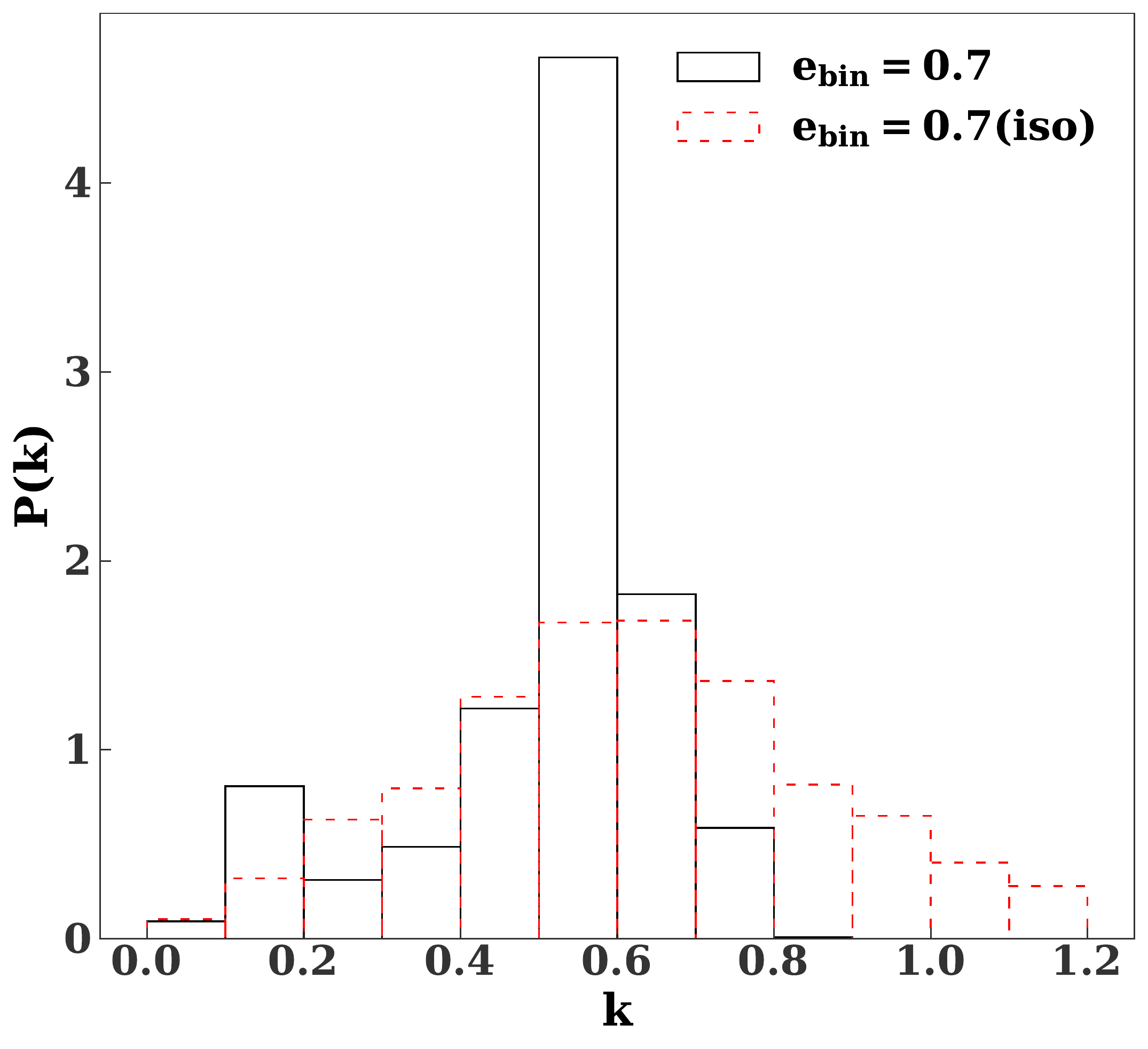}
    \caption{\label{fig:k} Left panel: probability distribution of $k$ (a dimensionless number that captures the scatter in the energy kick the binary components receive due to variations in phase and orientation) for binary eccentricities of $0$ (black, solid lines), and $0.7$ (red, dashed lines). In this panel the binary's internal angular momentum and the angular momentum of its centre-of-mass orbit are aligned. Right panel:
    Comparison of $k$ distributions for aligned (black, solid lines) and isotropic (red, dashed lines) binaries (with an eccentricity of 0.7). The distribution of energy kicks is not a strong function of binary inclination.}
\end{figure*}

We consider two different distributions for the binary inclination: (i) aligned with the orbital angular momentum and (ii) isotropic.
The left panel Figure~\ref{fig:k} shows the distribution of $k$ for aligned binaries with eccentricities of 0 and 0.7. To test the effects of binary inclination, we perform an additional set of simulations with an isotropic inclination distribution. As shown in the right panel of Figure~\ref{fig:k}, the $k$-distributions for isotropic and aligned binaries are similar. (However, the effective tidal radius is smaller for isotropic binaries--$\chi=1.7$ instead of 2.8.) Unless otherwise specified, we will present results for aligned binaries as in \citetalias{generozov&madigan2020}. However, the results for isotropic binaries are similar, as explicitly discussed in \S~\ref{sec:results}.

The binary semimajor axis is drawn from a log-uniform distribution. The minimum and maximum of this distribution are set by finite stellar radii and binary evaporation, respectively, and are computed as described in \citetalias{generozov&madigan2020}, except for the following improvements
\begin{enumerate}
    \item Unless otherwise noted, we use a smaller galactocentric radius for the binaries (0.05 pc\footnote{The inner edge in projection is $\approx 0.8''\approx 0.03$ pc (see \citealt{yelda+2014} and the references therein), which corresponds to a 3D distance $\approx 0.03-0.05$ pc, considering the orientation of the disk \citep{gillessen+2017}.} in lieu of 0.1 pc). The former is closer to the inner edge of the disk and would be more representative of disrupted binaries. (Though see discussion around Figure~\ref{fig:posteriora0}.)
    \item We use $5\times 10^5$ yr for the age of the binaries in lieu of $4\times 10^6$ yr. Although the disk is several Myr old \citep{lu+2013}, binary disruptions would generally occur early in the disk's history, and the former would be more representative of 
    disrupting binaries. (This change and the previous one roughly cancel out such that the maximum semimajor axis 
    is within a factor of a few of the prediction in \citetalias{generozov&madigan2020}.)
    \item We account for uncertainties in the disk mass function. Specifically, we adopt the \citet{lu+2013} measurement of the disk's mass function: $m^{-1.7\pm 0.2}$.\footnote{\citet{bartko+2010} claimed a much steeper mass function: $m^{-0.45\pm 0.3}$.} The number density, mean mass, and root-mean-square mass within the disk (which affect the evaporation time) are varied self-consistently with its mass function.
    The number of stars with masses $\geq 10 M_{\odot}$ is fixed to 280 for consistency with \citetalias{generozov&madigan2020}.
    \item We adjust the stellar radius of low-mass stars.
    The minimum binary semimajor axis is set by Roche-Lobe contact, and depends on stellar radius. In \citetalias{generozov&madigan2020} we assumed zero-age main sequence stellar radii. However, the timescale for low-mass stars to contract to the main sequence (i.e. the Kelvin--Helmholtz time) is significantly longer than the age of the young stars in the Galactic centre. We use the MIST \citep{dotter2016, choi+2016, paxton+2011,paxton+2013,paxton+2015} stellar isochrones\footnote{Version 1.2 isochrones with [Fe/H]=0 and $\Omega/\Omega_{\rm crit}=0.4$.} to evaluate stellar radii, assuming all (proto)stars are $5\times 10^5$ yr old.\footnote{This is the approximate timescale between the formation of the disk and the start of the binary disruption epoch \citep{generozov&madigan2020}.} At this time, stars with masses $\lsim 6.3 M_{\odot}$ are still contracting to the main sequence.
\end{enumerate}

The minimum of the disk (and binary) mass function is a free parameter, while the maximum mass is $100 M_{\odot}$. Such massive stars could be present at the epoch of binary disruption, though they would be removed over several Myr by stellar evolution.

In the following section, we compare the bound stars in these mock samples to the S-stars to infer constraints on the initial binary population in the clockwise disk.

\begin{table*}
\begin{threeparttable}
\caption{\label{tab:params} Distribution of binary parameters for mock disruptions.}
\begin{tabular*}{\textwidth}{@{\extracolsep{\fill}}lcccc}
    Property  & Distribution & Bounds & Parameters & Grid Points \\
    \hline
    Primary mass ($m_1$) & $m_1^{-n}$ & $m_{\rm min}\leq m_1\leq100 M_{\odot}$  & $n \in \{1.5, 1.6, 1.7, 1.8, 1.9\}$ & 5 \\
    {} & {} & {} & $m_{\rm min}\in \{0.5, 0.8, 1.2, 1.9, 3\}$ & 5\\
    \hline
    Mass ratio   ($q$)   &   $q^{-\alpha}$ &  $m_{\rm min}/m_1 \leq q \leq 1$             & $\alpha \in \{0, 0.1, ..., 2 \}$ & 21 \\
    \hline
    Eccentricity ($e_{\rm bin}$)   & $e_{\rm bin}^{\gamma}$ & $0 \leq e_{\rm bin}<1$                                               & $\gamma \in \{-0.5, 0, 1\}$ & 3\\
 \hline
\end{tabular*}
\begin{tablenotes}
 \item The primary mass, mass ratio, and eccentricity are sampled from the power-law distributions in the second column. The upper and lower bounds of these distributions are in the third column. Each distribution has $1-2$ free parameters (listed in the fourth column). We consider a grid for these parameters, with the values and dimensions indicated in the fourth and fifth columns.
\end{tablenotes}
\end{threeparttable}
\end{table*}

\section{Results}
\label{sec:results}
The observed S-stars have masses between $\sim 3$ and $\sim 15 M_{\odot}$ \citep{habibi+2017, cai+2018}. 
For each grid point in Table~\ref{tab:params}, we compute mock samples of $7\times 10^4$ disruptions, and calculate the probability that there would be no stars above $15 M_{\odot}$ among 19 massive stars ($m \geq 3 M_{\odot}$) drawn from each sample. This is the number of stars in Table 3 of \citet{gillessen+2017} that (i) have a semimajor axis within 0.03 pc, (ii) are not late-type stars, and (iii) are not members of the clockwise disk. We focus on smaller semimajor axes to avoid contamination from stars that may have been kicked out of the disk by other mechanisms (e.g. by vector resonant relaxation \citealt{szoelgyen&kocsis2018},  or by scattering between stellar mass black holes and disk binaries \citealt{trani+2019}).  \agnote{We implicitly assume these stars all belong to a single, few Myr-old population, although some admixture of older stars is possible considering (i) only 8 of the 19 stars have spectroscopic age measurements (ii) spectroscopy of recently-discovered, faint S-stars indicate ages closer to $\sim 100$ Myr \citep{peissker+2020}.}

The most massive stars in each sample would have died off over the $\sim$5 Myr since the epoch of disk formation and binary disruption. The largest plausible present-day mass depends on stellar metallicity and rotation as shown in Table~\ref{tab:mto}. Here, we remove any stars with (initial) masses above $38 M_{\odot}$, the approximate main sequence turnoff mass for a 5 Myr old, solar metallicity population.

\begin{table}
\begin{threeparttable}
\caption{\label{tab:mto} Dependence of the main sequence turnoff mass (at 5 Myr) on stellar metallicity and rotation.}
\begin{tabular*}{\columnwidth}{@{\extracolsep{\fill}}lcc}
 [Fe/H]  & $M_{\rm to} [M_{\odot}]$  &  $M_{\rm to} [M_{\odot}]$   \\
 \hline
   & $\Omega/\Omega_{\rm crit}=0$ & $\Omega/\Omega_{\rm crit}=0.4$\\
 \hline
        -2     &        40.6 (40.3)     &  50.0 (48.8)\\
        0      &        37.1 (33.8)     &  40.0 (34.9)\\
        0.5    &        35.9 (18.5)     &  35.1 (26.4)\\
 \hline
 \end{tabular*}
 \begin{tablenotes}
 \item Second column is the turnoff masses for non-rotating stars, while the third column corresponds to the turnoff mass for rotating stars with a (zero-age main sequence) rotational velocity of $\Omega=0.4 \Omega_{\rm crit}=0.4 \left[(1-L_*/L_{\rm edd}) G m_\star/r_\star^3\right]^{1/2}$, where $m_\star$, $r_\star$, and $L_\star$ are the stellar mass radius and luminosity; $L_{\rm edd}$ is the Eddington luminosity. The first number in each cell is the initial mass, while the second number in parentheses is the final mass after mass loss due to winds.
 \end{tablenotes}
\end{threeparttable}
\end{table}

\begin{figure*}
        \includegraphics[width=\textwidth]{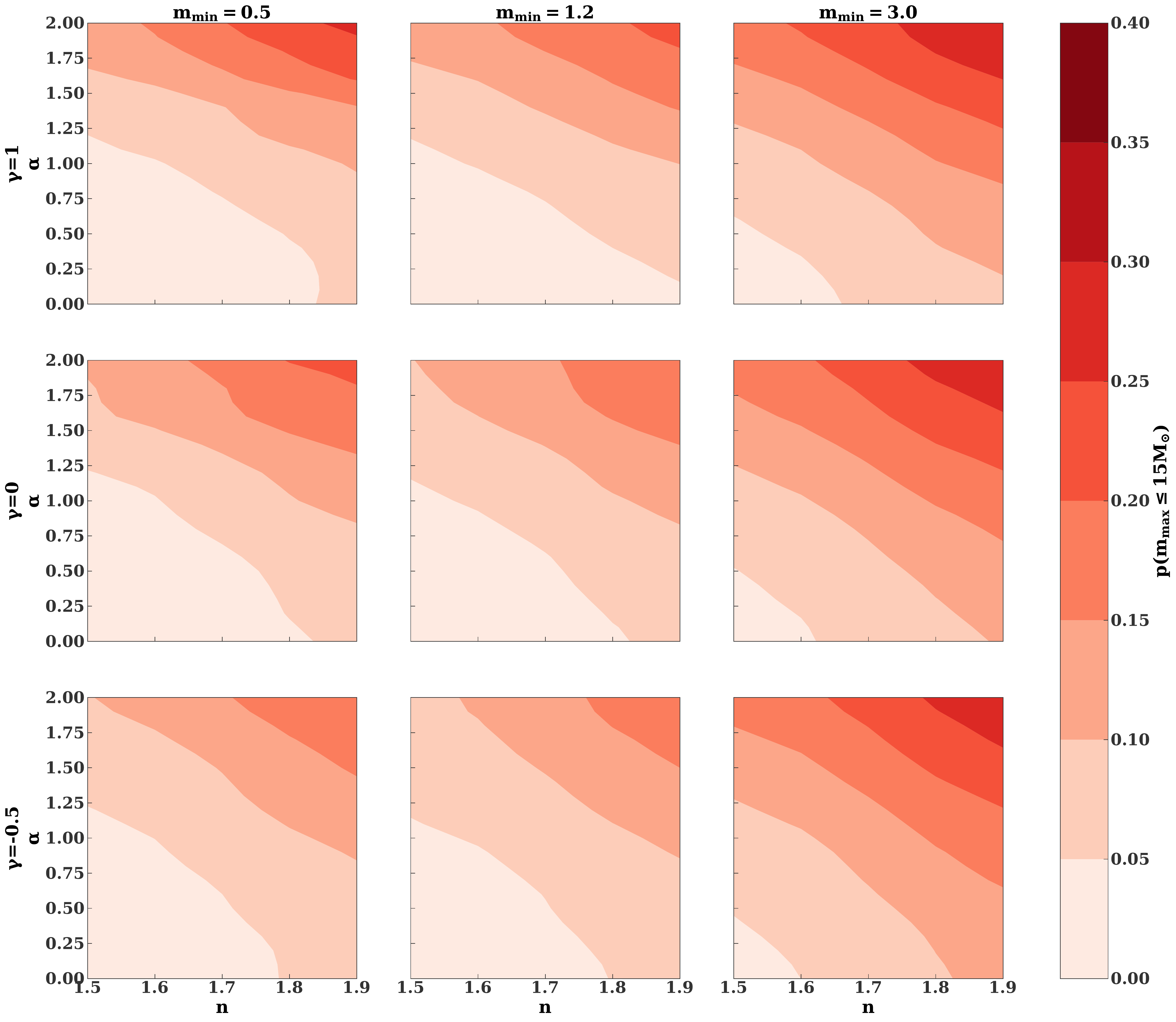}
        \caption{\label{fig:posterior} Probability of having no S-stars above $15 M_{\odot}$ for different binary parameters. Each column corresponds to a minimum stellar mass ($m_{\rm min}$ in Table~\ref{tab:params}), and each 
        row corresponds to an eccentricity distribution slope ($\gamma$). Within each cell the probability is plotted as a function of the slopes of the mass ratio and primary mass distributions ($\alpha$ and $n$ in Table~\ref{tab:params} respectively).}
    \end{figure*}

\begin{figure*}
        \includegraphics[width=\textwidth]{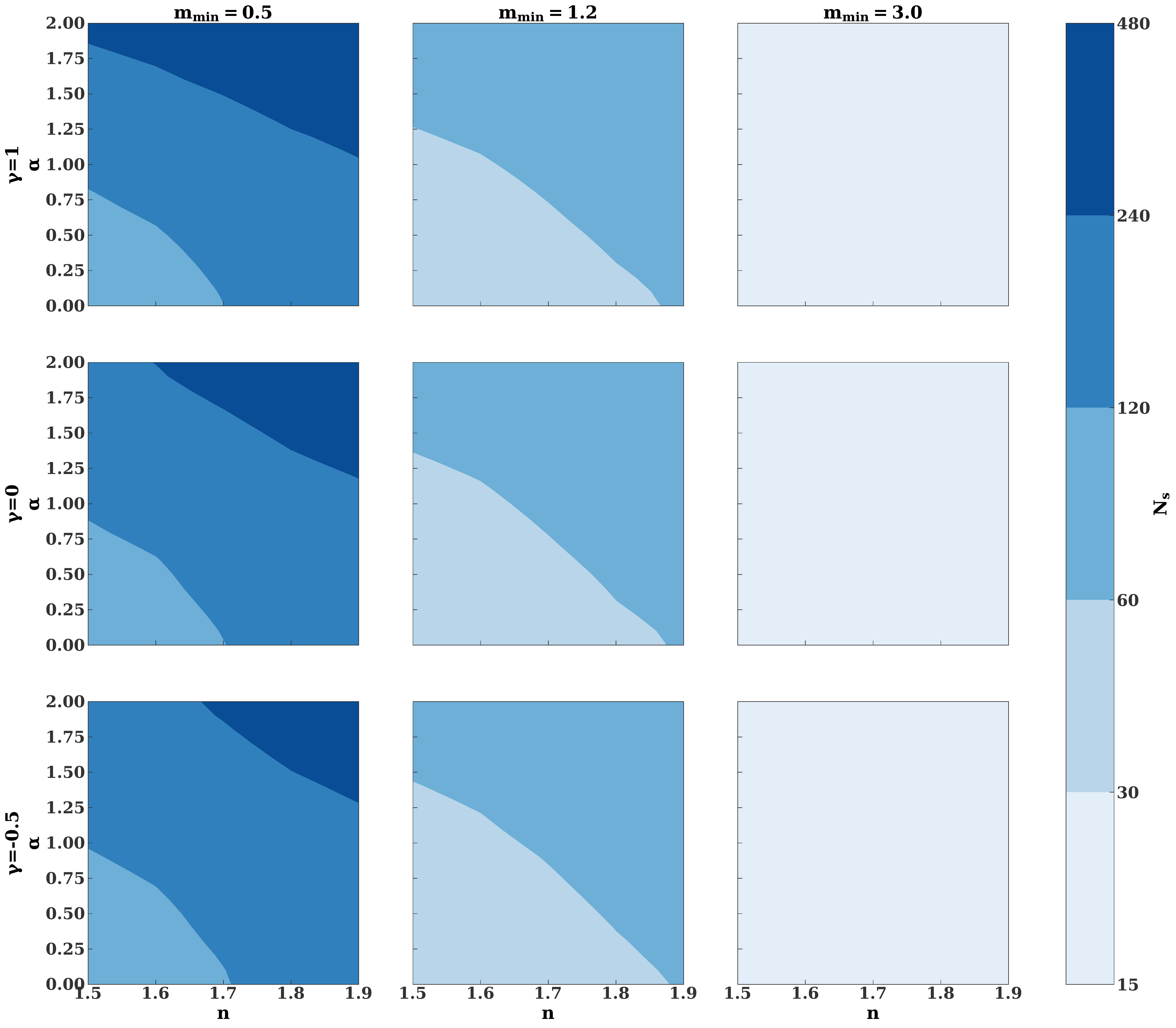}
        \caption{\label{fig:number} Total number of S-stars ($N_s$) originating from disruptions of disk binaries across the parameter space in Figure~\ref{fig:posterior}. In other words, this is the total number of disruptions necessary to produce the 19 observed stars. The hundreds of stars in the left-most column are mostly lower mass ($<3 M_{\odot}$), and would not be among the observed population.}
    \end{figure*}

Stellar winds would reduce the maximum mass further. However, winds only have a significant effect on the largest surviving stars. (In 5 Myr, a $38 M_{\odot}$ solar metallicity star is reduced to $33.7 M_{\odot}$, while a $20 M_{\odot}$ star is reduced to $19.8 M_{\odot}$.)
Therefore, stellar wind mass loss has a negligible effect on the probability of no S-stars above $15 M_{\odot}$. 
Figure~\ref{fig:posterior} shows this probability as a function of $\alpha$, $n$, $\gamma$, the power law slopes of the mass ratio, primary mass, and eccentricity distributions, respectively, as well as $m_{\rm min}$, the minimum mass of the disk mass function. To reduce numerical noise, we have applied a Gaussian filter with a standard deviation of 1 grid point to the probability data.\footnote{ \texttt{scipy.ndimage.gaussian\_filter(data,1)}).} Increasing $\alpha$ favors lower mass secondary stars, which accentuates the mass ratio effect described in $\S$~\ref{sec:mr}. This increases the probability of no stars above $15 M_{\odot}$, $p(m_{\rm max} \leq 15 M_{\odot})$. A bottom-heavy mass function (larger $n$) also results in a larger probability. The probabilities are not a strong function of the minimum mass in the disk, but the total number of S-stars originating from disruption of disk binaries is. The total number of such stars varies from 19 to 460 across our parameter space (see Figure~\ref{fig:number}). In the latter case, most of these stars would be low-mass and unobserved.

\begin{figure*}
        \includegraphics[width=\columnwidth]{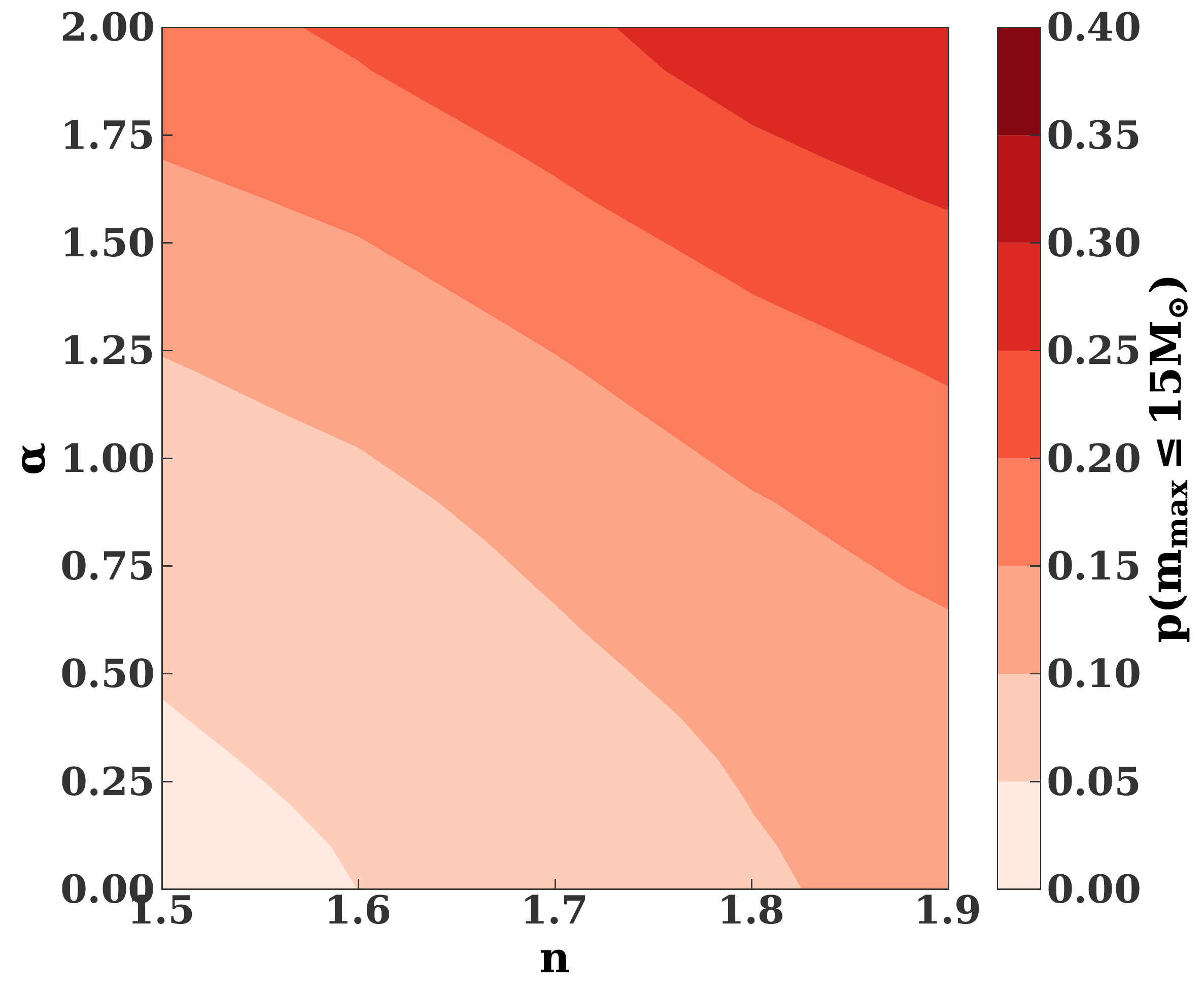}
        \includegraphics[width=\columnwidth]{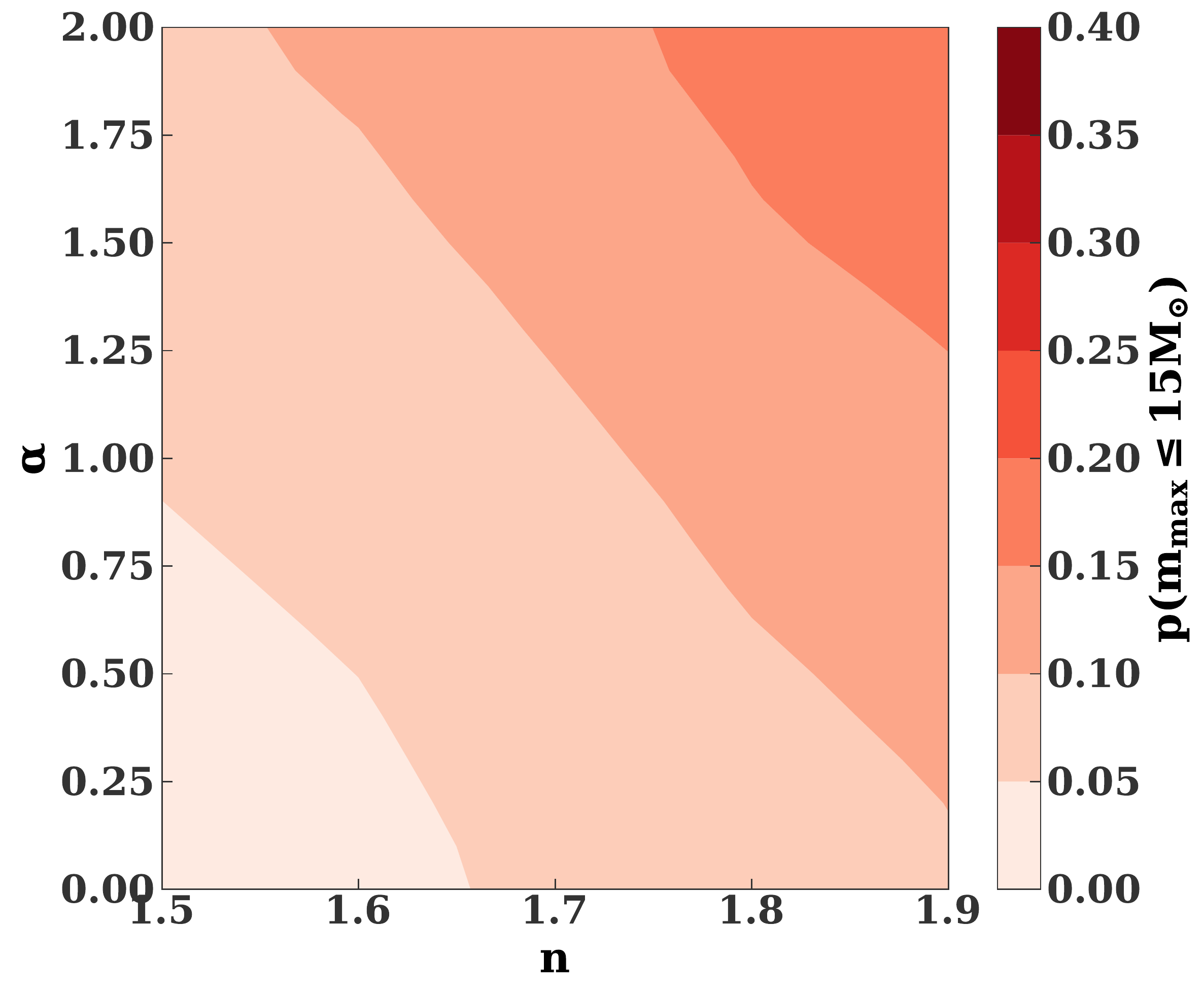}
        \caption{\label{fig:posteriorMarg} Left panel: probability of no S-stars above $15 M_{\odot}$ as a function of the slopes of the mass ratio distribution, $\alpha$, and the primary mass function, $n$. The probability is maximized over all other parameters. Right panel: probability of no S-stars above $15 M_{\odot}$, assuming random sampling from binary stars. This panel shows the contribution of sampling effects to the results.}
\end{figure*}

\agnote{Figure~\ref{fig:posteriorMarg} shows the maximum of $p(m_{\rm max} \leq 15 M_{\odot})$ over all eccentricity distributions and minimum masses as a function of $\alpha$ and $n$. For every mass function slope, $n$, we compute the minimum mass ratio slope, $\alpha$, where probability is greater than or equal 10\%. Based on a linear fit to this data, the probability is at least 10\% for}
\begin{equation}
    \alpha \gsim \alpha_c = -3.5 (n-1.7)+0.6.
\end{equation}
\agnote{Considering, the steep dependence of $\alpha_c$ on $n$, a broad range of mass ratio slopes are plausible including those from \citet{kiminki&kobulnicky2012} and \citet{sana+2013}. The right panel of Figure~\ref{fig:posteriorMarg} shows $p(m_{\rm max} \leq 15 M_{\odot})$, assuming the S-stars are randomly sampled from the binaries in our Monte-Carlo ensembles. This shows the contribution of sampling effects (as opposed to binary disruption physics) to the probability. While the probabilities with only sampling effects are smaller, they are non-negligible for bottom-heavy mass functions.}

So far, we have assumed the internal binary orbit and the centre-of-mass orbit are aligned. (This assumption affects the ``k'' distribution and the effective tidal radius--see \S~\ref{sec:mc}). In fact, isotropic binaries would give similar results. (For example, the probabilities in Figure~\ref{fig:posterior} would be $\sim 0.9-1.3$ times as large with isotropic binaries. For isotropic binaries,  probabilities would be higher (lower) for small (large) $\alpha$.)  

\begin{figure*}
        \includegraphics[width=\textwidth]{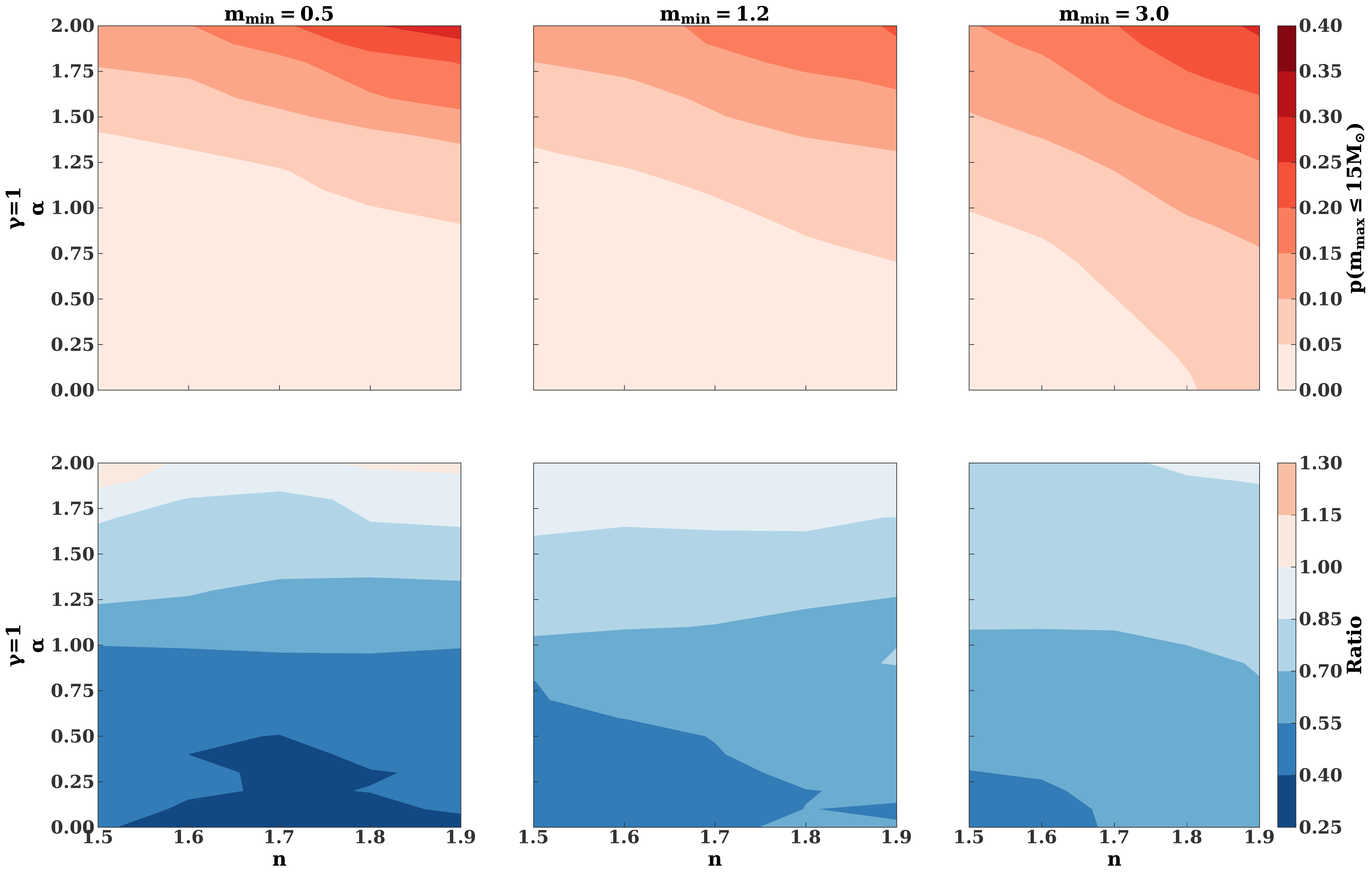} 
        \caption{\label{fig:posteriora0} Effect of the binary center-of-mass orbit's semimajor axis on the probability of no S-stars above $15 M_{\odot}$. Top panel: same as the top row of Fig.~\ref{fig:posterior}, except the semimajor axis of the center-of-mass is 0.1 pc, instead of 0.05 pc. Bottom panel: ratio between the probabilities for semimajor axes of 0.1 pc and 0.05 pc. Increasing the semimajor axis generally decreases the probability of no massive S-stars.}
    \end{figure*}

We now briefly discuss the effect of the initial binary semimajor axis on the results. So far we have fixed this semimajor axis to 0.05 pc, near the present-day inner edge of the disk. In the eccentric disk simulations of GM20, particles near the inner edge are preferentially disrupted, but there is still a factor of a few spread in the semimajor axes of disrupting particles. Additionally, this spread will depend on the semimajor axis distribution of the disk (In GM20, the semimajor axis distribution was proportional to $a^{-1}$). Finally, the inner edge of the disk may have been larger in the past, considering binary disruptions could shift the inner edge of the disk to larger semimajor axes. Therefore, we explore how shifting the initial semimajor axis of the binaries to 0.1 pc affects the results. (We also adjust the internal binary semimajor axis distribution, accounting for a larger galactocentric radius; see \S~\ref{sec:mc}.) Figure~\ref{fig:posteriora0}, shows how the probabilities in the top row of Figure~\ref{fig:posterior} change with this shift. The results are qualitatively similar, but the probabilities are generally smaller. In this case, the probability of no massive S-stars is at least ten percent for
\begin{equation}
    \alpha \gsim \alpha_c = -2.6 (n-1.7)+1.0
\end{equation}

\begin{figure*}
        \includegraphics[width=\textwidth]{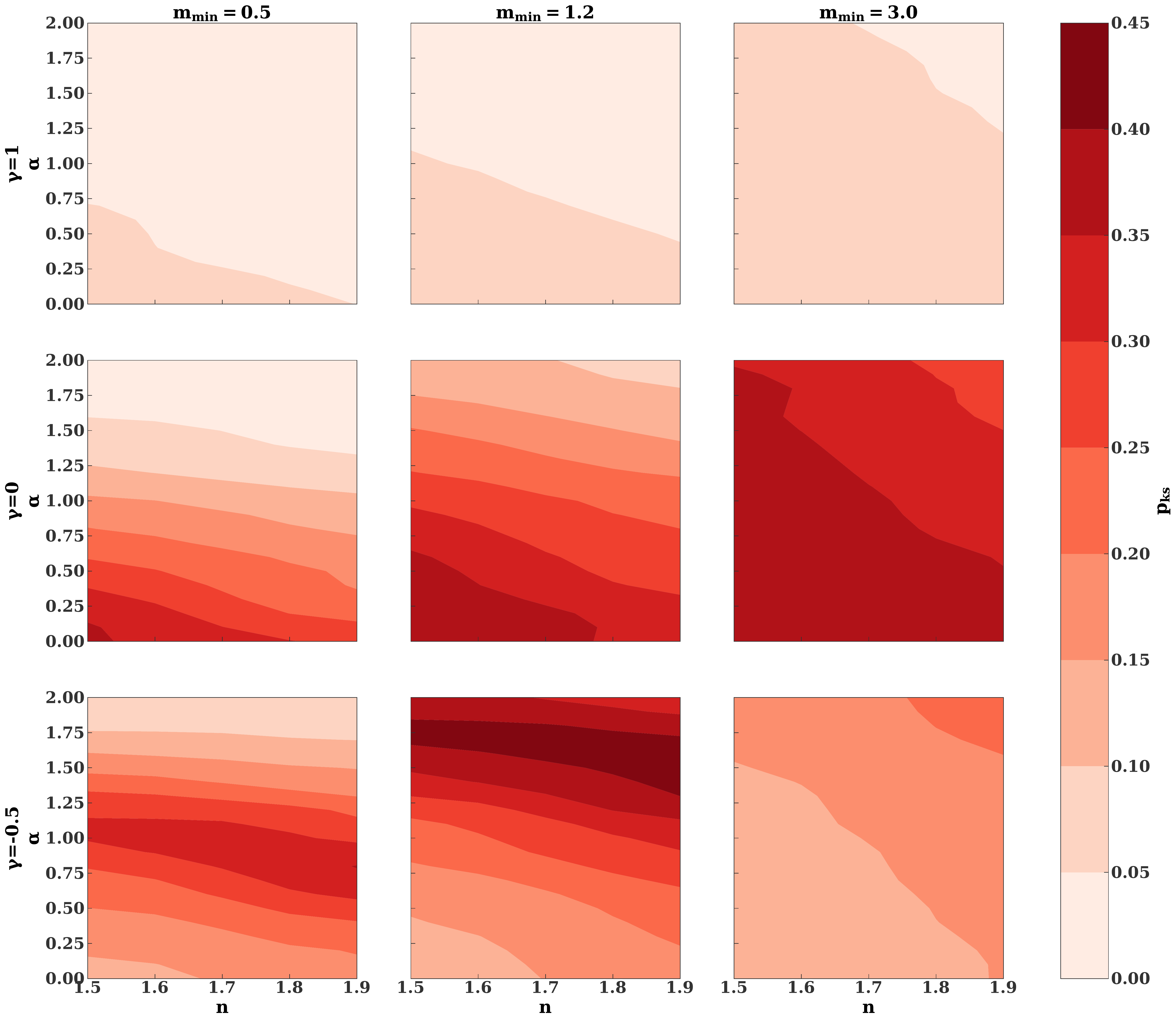}
        \caption{\label{fig:ks} Two sample K--S test p-values that the simulated and observed S-star semimajor axis distributions are consistent, as a function of binary parameters (e.g. the primary mass, eccentricity, and mass ratio distributions). The layout is the same as Figure~\ref{fig:posterior}.}
    \end{figure*}

Our model should also reproduce the semimajor axis distribution of the S-stars, considering two-body relaxation is unlikely to significantly perturb the S-stars' semimajor axes over their lifetimes (see the review by \citealt{talalexander2017}).\footnote{However, the S-stars' inclination and eccentricity distributions can be significantly perturbed over several Myr by resonant relaxation \citep{rauch&tremaine1996, kocsis&tremaine2011, antonini&merritt2013, bar-or&fouvry2018, generozov&madigan2020}.} Figure~\ref{fig:ks} shows the p-values from two-sample K--S tests applied to the observed S-stars and the mock binary samples in Figure~\ref{fig:posterior}. At face value, this statistic excludes significant portions of parameter space. However, caution is warranted in interpreting this test, since (i) the observed distribution is likely affected by position dependent completeness effects and (ii) some of the stars deposited by binary disruptions can diffuse into the loss cone and be tidally disrupted \citep{wenbinlu+2020}.

Could the three known binaries in the clockwise disk be used to constrain the mass ratio distribution? The known binaries have mass ratios $\gsim 0.5$. The probability of drawing three such binaries from a $q^{-\alpha}$ distribution extending from 0.05\footnote{The primary mass in these binaries is $\gsim 20 M_{\odot}$, which gives a minimum mass ratio of $\sim$0.05 if the mass function extends to $1 M_{\odot}$.} to 1 is less than 5\% for $\alpha>0.53$. However, this analysis assumes that the observed binaries are an unbiased sample of the binary population in the disk several Myr ago. This is probably not the case, considering two of them (E60 and 1RS16SW) are contact binaries \citep{martins+2006, pfuhl+2014}, that may be affected by mass transfer.

Binary disruption also ejects stars from the Galactic center. In fact, \citet{generozov2020} predicts that the Galactic centre produced a conical stream of high velocity stars $\sim$5 Myr ago. The properties of this stream depend on the properties of the progenitor binaries. For example, a bottom-heavy mass ratio distribution translates into a bottom-heavy stream mass function. For $\alpha=0$, the power-law slope of the stream mass function is comparable to the slope of the disk mass function ($\sim 1.7$). For $\alpha=2$, this slope is significantly steeper ($\sim 2.4-2.7$).
Additionally, the number of stream stars will vary in tandem with the total number of disruptions (see Figure~\ref{fig:number}). For $m_{\rm min}=3 M_{\odot}$ there would be less than 10 high velocity stars, while for $m_{\rm min}=0.5 M_{\odot}$ there would be $\sim 30-100$. (Note that as in \citet{generozov2020}, we consider stars with velocities above 500 km s$^{-1}$ to be high velocity.)

\section{Stellar Disruptions}
\label{sec:dis}
Some of the stars that were originally deposited in the S-star cluster may have been tidally disrupted \citep{wenbinlu+2020}. Massive stars may be preferentially disrupted due to their larger tidal radii, especially at later evolutionary stages \citep{chen&amaro-seoane2014}. Here, we estimate the fraction of stars that would be disrupted as a function of mass by performing Monte-Carlo simulations of angular momentum diffusion due to two-body relaxation as described in \citet{wenbinlu+2020}. This work found that the former companion of the hypervelocity star S5-HVS1 would have an $\sim 80-90\%$ probability for disruption via tidal heating for a two-body relaxation time of $\sim 10^8$ years. In the adiabatic limit (as assumed in \citealt{wenbinlu+2020}), a star will be disrupted over $N$ orbits if 

\begin{equation}
    \sum_{i=1}^N \frac{\delta E_i}{G m_\star^2/r_{\star, i}} \geq 0.1,
    \label{eq:cond1}
\end{equation}
where $\delta E_i$ is the tidal heating on its ``ith'' orbit around the central SMBH; $m_\star$ and $r_\star$ are the stellar mass and radius respectively. 

S5-HVS1 had an unusually high velocity, and its former companion would have a correspondingly small semimajor axis \citep{koposov+2019}. In fact the disruption probability would be significantly smaller at larger semimajor axes since (i) stars at larger semimajor axes complete less orbits in a given time interval and (ii) the loss cone will be smaller. 

We calculate the disruption probability near the inner and outer edges of the S-star cluster (0.005 and 0.03 pc respectively) for five discrete stellar masses and two angular momenta. In particular, for each row in Table~\ref{tab:dis} we simulated $10^4$ random walks in angular momentum (at fixed semimajor axis, $a_s$), using the procedure described in \S~4.2 in \citet{wenbinlu+2020}. The two-body relaxation time is
\begin{equation}
    t_{\rm rx}=1.6\times 10^8 {\rm yr} \left(\frac{a_s}{0.005 \,\,{\rm pc}}\right)^{0.25},
\end{equation}
as motivated by previous work on collisional relaxation in the Galactic centre \citep{antonini&merritt2013, generozov+2018}.\footnote{The two-body relaxation time between $10^{-2}$ and $10^{-3}$ pc in the former (latter) reference is $\sim 1.6\times 10^8 \left(r/0.005 {\rm pc}\right)^{0.46}$ ($\sim 2.7\times 10^8 \left(r/0.005 {\rm pc}\right)^{0.25}$) yr.} Although resonant relaxation would also affect the angular momentum distribution of the S-stars \citep{rauch&tremaine1996, antonini&merritt2013}, it would not strongly affect the stellar disruption rate, because resonant relaxation is suppressed by general relativistic precession at low angular momenta \citep{madigan+2011,merritt+2011}. 

The fourth column of Table~\ref{tab:dis} shows the disruption probability assuming equation~\eqref{eq:cond1} for the disruption criterion and 
\begin{align}
    &\frac{\delta E_i}{G m_\star^2/r_\star} = T_2\left(\frac{r_p}{r_t^\star},m_\star,t\right) \left( \frac{r_p(t)}{r_t^\star(m_\star, t)}\right)^{-6} \nonumber\\
    &r_t^\star(m_\star, t)=\left(\frac{M}{m_\star}\right)^{1/3} r_\star(m_\star, t),
\end{align}
for the tidal heating per orbit.
Here $T_2(\eta)$ is the quadrupole tidal coupling constant, which is calculated as described in Appendix~\ref{app:tides}; $r_t^\star$ is the star's tidal radius and $r_p$ is the pericenter of its orbit. For this table, we assume stars start on the zero-age main sequence. 
Overall, the disruption probability ranges from 2\% to 60\% in Table~\ref{tab:dis}, depending on the star's semimajor axis, initial angular momentum, and mass.

\begin{table}
\begin{threeparttable}
    \centering
    \caption{\label{tab:dis} Stellar disruption probabilities as a function of stellar mass, (initial) angular momentum, and semimajor axis.}
    \label{tab:my_label}
    \begin{tabular*}{\columnwidth}{@{\extracolsep{\fill}}lccccc}
    $M_*$ [$M_{\odot}$] & $j/j_c$  & $a_c$ & $p_d$ & $p_{d, 2 r_t}$ & $p_{d, r_t}$\\
    \hline
              40 &    0.26 &  0.03 &        0.042 &          0.067 & 0.045 \\
              40 &    0.16 &  0.03 &         0.18 &           0.24 &  0.18 \\
              40 &    0.26 & 0.005 &         0.24 &            0.3 &  0.19 \\
              40 &    0.16 & 0.005 &         0.56 &           0.62 &  0.45 \\
              20 &    0.26 &  0.03 &         0.03 &          0.039 & 0.028 \\
              20 &    0.16 &  0.03 &         0.15 &           0.18 &  0.14 \\
              20 &    0.26 & 0.005 &         0.18 &           0.19 &  0.13 \\
              20 &    0.16 & 0.005 &         0.47 &           0.49 &  0.37 \\
              10 &    0.26 &  0.03 &        0.027 &          0.035 & 0.025 \\
              10 &    0.16 &  0.03 &         0.14 &           0.16 &  0.13 \\
              10 &    0.26 & 0.005 &         0.14 &           0.15 &  0.11 \\
              10 &    0.16 & 0.005 &         0.41 &           0.43 &  0.32 \\
                5 &    0.26 &  0.03 &        0.025 &          0.032 & 0.025 \\
                5 &    0.16 &  0.03 &         0.13 &           0.15 &  0.12 \\
                5 &    0.26 & 0.005 &         0.12 &           0.14 & 0.095 \\
                5 &    0.16 & 0.005 &         0.37 &            0.4 &  0.29 \\
                2 &    0.26 &  0.03 &        0.021 &          0.026 & 0.021 \\
                2 &    0.16 &  0.03 &         0.11 &           0.14 &  0.11 \\
                2 &    0.26 & 0.005 &         0.11 &           0.12 & 0.088 \\
                2 &    0.16 & 0.005 &         0.33 &           0.36 &  0.28 \\
                1 &    0.26 &  0.03 &        0.021 &          0.024 &  0.02 \\
                1 &    0.16 &  0.03 &         0.11 &           0.12 & 0.099 \\
                1 &    0.26 & 0.005 &          0.1 &            0.1 & 0.077 \\
                1 &    0.16 & 0.005 &         0.31 &           0.31 &  0.24 \\
    \hline 
    \end{tabular*}
\begin{tablenotes}
\item The disruption probability in the fourth column is computed using equations~\eqref{eq:cond1} for the disruption criterion. The probabilities in the last two columns are computed assuming equation~\eqref{eq:cond2} with  $\lambda$ as indicated in the column header's subscript.
\end{tablenotes}
\end{threeparttable}
\end{table}

As previously discussed, stars with masses $\lsim$ a few $M_{\odot}$ would not have reached the main sequence within several Myr, thus the results in Table~\ref{tab:dis} are not directly applicable to the S-stars. 
However, this table suggest a simple prescription for disruption, which can also be used for protostars.
The last two columns of this table show the disruption probability, assuming stars are disrupted when their pericenter satisfies
\begin{equation}
    r_p\leq r_{p,c}=\lambda \left(\frac{M}{m_\star}\right)^{1/3} r_\star,
    \label{eq:cond2}
\end{equation}
where $\lambda$ is either 1 (last column) or 2 (second-to-last column). Overall equation~\eqref{eq:cond2} gives disruption probabilities comparable to those found with the more precise tidal heating calculation.

Considering that tidal heating scales as $r_p^{-6}$, we expect $r_{p,c}\propto N_{\rm orb}^{1/6} \propto a_s^{-1/4}$, where $N_{\rm orb}$  is the number of stellar orbits and $a_s$ is the semimajor axis. Based on Table~\ref{tab:dis}, a reasonable approximation for $\lambda$ is 
\begin{equation}
    \lambda \approx 2 \left(\frac{a_s}{0.03 {\rm pc}}\right)^{-1/4}
    \label{eq:lam}
\end{equation}
for S-star semimajor axes $a_s\leq 0.03$ pc. Thus, we can track whether stars are disrupted using equations~\eqref{eq:cond2} and~\eqref{eq:lam}, with the appropriate stellar radii from the MIST stellar evolution tracks \citep{dotter2016, choi+2016, paxton+2011, paxton+2013, paxton+2015}.\footnote{Version 1.2 isochrones with [Fe/H]=0 and $v/v_{\rm crit}=0.4$.}

\begin{figure*}
        \includegraphics[width=\textwidth]{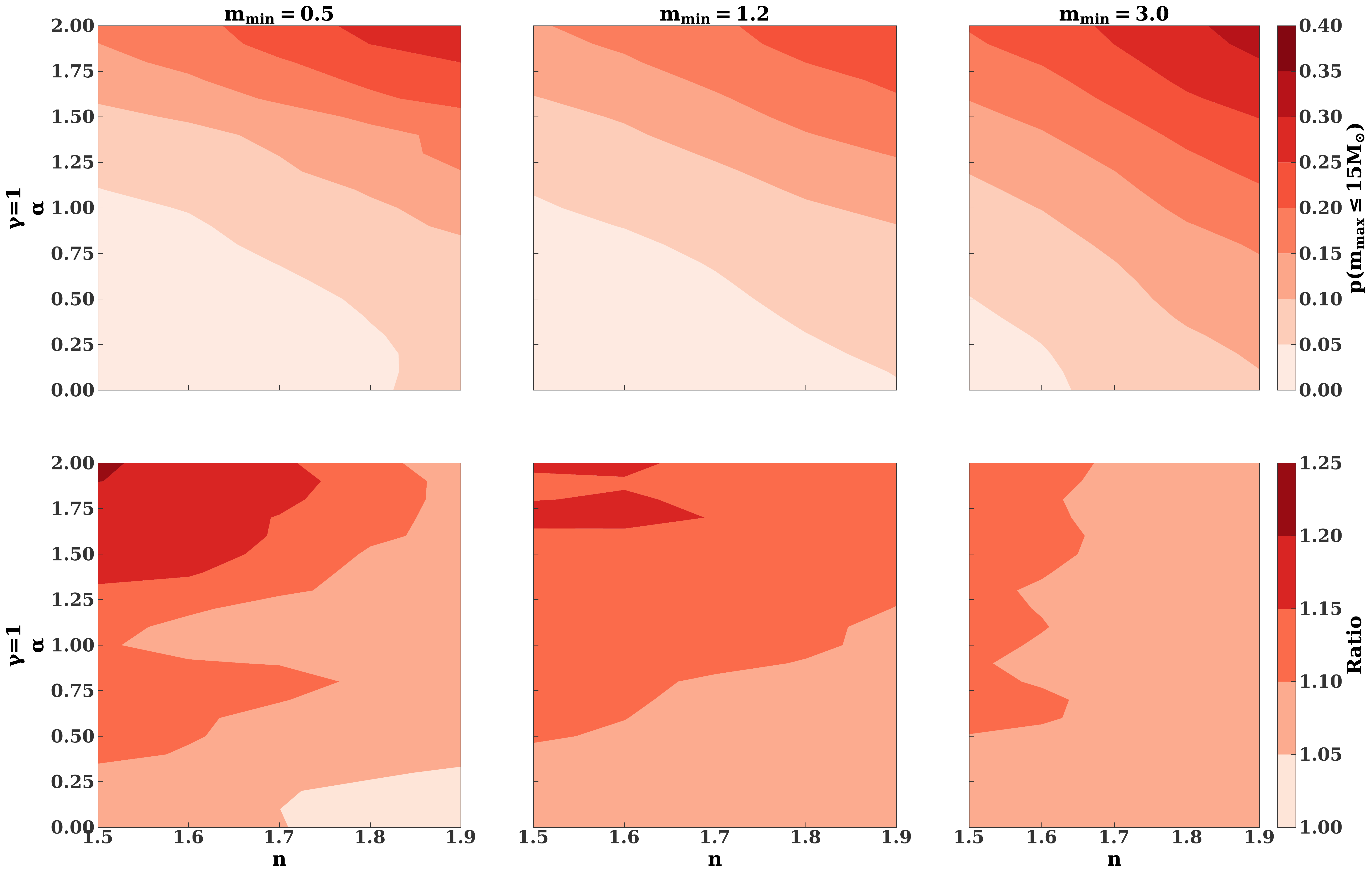} 
        \caption{\label{fig:posteriordis} Top panel: probability of no S-stars above $15 M_{\odot}$ after accounting for stellar disruptions (as described in \S~\ref{sec:dis}) in the top row of Figure~\ref{fig:posterior}. Bottom panel: Ratio of post-stellar disruption probabilities to pre-stellar disruption probabilities.}
    \end{figure*}

We simulate a random walk in angular momentum for all of the stars in the mock S-star samples in \S~\ref{sec:mc}. All stars are initially $5\times 10^5$ yr old, and random walk for 4.5 Myr. We delete stars that are disrupted according to equations~\eqref{eq:cond2} and~\eqref{eq:lam}. 
Figure~\ref{fig:posteriordis} shows the effect of deleting such stars from the top row Figure~\ref{fig:posterior}. The probability of no stars above $\sim 15 M_{\odot}$ can change by a factor of $\sim 1.2$. Overall, stellar disruptions do not strongly affect the mass distribution of the S-stars because (i) disruptions are fairly rare at larger semimajor axes and (ii) for small $\alpha$ there is a positive correlation between the initial angular momentum and stellar mass, which decreases the disruption probability of massive stars. (For equal mass binaries we expect $1-e_s\approx \left(m_{\rm bin}/M\right)^{1/3}$, where $e_s$ is the S-star eccentricity. Thus, the eccentricity decreases with binary mass; \citetalias{generozov&madigan2020}.)

\section{Summary}
\label{sec:conc}
Observations suggest that many of the Galactic centre S-stars come from disruption of binaries from a surrounding disk structure. However, unlike the disk, the S-stars do not include any stars above $15 M_{\odot}$. In this paper, we show this discrepancy could be due to a combination of sampling effects and a built-in bias in the binary disruption process. Namely, the primary star remains closer in energy to the original centre-of-mass orbit than the secondary star. Bottom-heavy mass functions and mass ratio distributions increase the chances that the S-star cluster is not populated by massive stars. For binary disruptions to explain the S-star cluster they must inject approximately 19 stars with masses above 3 $M_{\odot}$ inside of 0.03 pc. With this constraint, there is a $>10\%$  chance that binary disruptions would not leave any stars above $15 M_{\odot}$ in the S-star cluster, as long as the power-law slope of the binary mass ratio distribution satisfies
\begin{equation}
    \alpha\gsim -3.5 (n-1.7) +0.6,
\end{equation}
where $n$ is the power-law slope of the primary mass function.

Stellar disruptions have a minor impact on the results. In principle, high mass stars are more susceptible to disruptions. In practice, this is a weak effect, considering a realistic initial distribution of S-star angular momenta and semimajor axes.

\section*{ACKNOWLEDGEMENTS}
{

I thank the anonymous referee for a constructive report.

I thank Ann-Marie Madigan for insightful feedback, as well as continuous support and encouragement on this project. I thank Jason Dexter, Brian Metzger, and Nicholas Stone for helpful conversations.

I thank Adam Litzler and Christopher Zimmer from the Laboratory for Interdisciplinary Statistical Analysis (LISA) at the University of Colorado Boulder for their feedback on the data summarization and visualization.

I acknowledge support from NASA Astrophysics Theory Program (ATP) under grant NNX17AK44G.

This work utilized resources from the University of Colorado Boulder Research Computing Group, which is supported by the National Science Foundation (awards ACI-1532235 and ACI-1532236), the University of Colorado Boulder, and Colorado State University. 
}

\textit{Software:} \texttt{AR--Chain} \citep{mikkola&merritt2008}, \texttt{REBOUND} \citep{rein.liu2012}, 
\texttt{REBOUNDX} \citep{tamayo+2019},
\texttt{AstroPy} \citep{astropy+2018}, \texttt{Matplotlib} \citep{hunter+2007}, \texttt{NumPy}, \texttt{SciPy} \citep{2020SciPy-NMeth}, IPython \citep{perez+2007}, \texttt{MESA} \citep{paxton+2011, paxton+2013, paxton+2015, paxton+2018, paxton+2019}, \texttt{MIST} \citep{dotter2016, choi+2016}

\section*{Data Availability}

Inlists and other scripts necessary for the stellar evolution and mode analysis calculations in Appendix~\ref{app:tides} are available at \doi{10.5281/zenodo.4081615}. Other  data  will  be  made  available upon reasonable request.

\appendix
\section{Tidal heating}
\label{app:tides}
This appendix describes how the tidal coupling constants in \S~\ref{sec:dis} are calculated. The calculation can be broken down into a couple of steps
\begin{enumerate}
\item For each stellar mass in Table~\ref{tab:dis}, we evolve a MESA\footnote{Version 12778; SDK x86\_64-linux-20.3.2} \citep{paxton+2011, paxton+2013, paxton+2015, paxton+2018, paxton+2019}
model for 5 Myr, starting from the zero-age main sequence. The relevant inlists have been uploaded to Zenodo (\doi{10.5281/zenodo.4081615}). Stellar rotation and mass loss are neglected in the calculation.
\item We output pulsation profiles from this calculation, and input them into the GYRE\footnote{Version 5.0}
stellar oscillation code \citep{townsend+2013}. The inlists for the GYRE calculation are also available on Zenodo (\doi{10.5281/zenodo.4081615}). The frequency grid, for each mass and snapshot are generated automatically via an included python script.
The GYRE calculation gives the normal mode spectrum of each star. 
\end{enumerate}
Stellar evolution is only important for stars with masses $\gsim 20 M_{\odot}$ on the timescales of interest. For lower mass stars, the zero-age main sequence tidal coupling constants would be an adequate approximation.

The quadrupole tidal coupling constant is 
\begin{equation}
    T_{2}(r_p/r_t, m_\star, t)=  \sum_{\alpha, m} 2 \pi^2 Q_{\alpha, \ell=2}^2 K_{\alpha, \ell=2, m}^2,
    \label{eq:tcoup}
\end{equation}
where $Q$ and $K$ are the overlap integrals defined in \citet{press&teukolsky1977}, and the summation is over mode number ($\alpha$) and azimuthal degree ($m$); $Q$ can be evaluated via    

\begin{align}
    &Q_{\alpha, \ell}=\frac{(2 \ell+1) \delta \phi_\star}{4 \pi I^{1/2}}\nonumber\\
    &I_{\alpha}=\int_0^1 x^2 \tilde{\rho} (\xi_{r, \alpha}^2 +\ell (\ell+l) \xi _{h, \alpha}^2) dx,
    \label{eq:overlap}
\end{align}
where $\delta \phi_\star$ is the (Eulerian) potential perturbation at the stellar surface in units of $G m_\star/r_\star$ (see equation 9 in \citealt{burkart+2012}); $I$ is an overall normalization; $\xi_{r, \alpha}$ and $\xi_{h, \alpha}$ are the radial and tangential displacement eigenfunctions (in units of stellar radius); $\tilde{\rho}$ is the stellar density (in units of $m_\star/r_\star^3$); $x$ is a dimensionless radial coordinate. GYRE computes $\delta \phi_\star$, $\xi_r$, and $\xi_h$, while $\tilde{\rho}$ comes from the underlying MESA models.

In general, the summation in equation~\eqref{eq:tcoup} is dominated by the fundamental mode and $g$-modes. High frequency, low-order modes dominate at smaller pericenters, while low frequency, high-order modes dominate at large pericenters. In general, we can only compute a finite set of modes, and the summation becomes inaccurate for pericenters
\begin{equation}
    \frac{r_{p}}{r_t} \gsim 2 \tilde{\omega}(\alpha_{\rm min})^{-2/3}
    \label{eq:rpAcc}
\end{equation}
where $\alpha_{\rm min}$\footnote{By convention g-modes are denoted by negative mode numbers.} and $\tilde{\omega}(\alpha_{\rm min})$ are the order and frequency of the highest-order g-mode included (in units of the characteristic stellar frequency, $\left(G m_\star/r_\star^3 \right)^{1/2}$). We find that $\alpha_{\rm min}=-30$ is sufficient for pericenters $r_p/r_t\leq 4.6.$\footnote{The inequality in equation~\eqref{eq:rpAcc} is satisfied for all stars in Table~\ref{tab:dis}, except for $40 M_{\odot}$ ones. These are treated with the procedure described in the next paragraph} We use a power-law extrapolation for larger pericenters. The details of this extrapolation are not important, as the tidal heating from larger pericenters is negligible. At this pericenter the tidal coupling constant is $\lsim 10^{-4}$, so that $\lsim 10^{-8}$ times the stellar binding energy is deposited into the star per orbit. Over 5 Myr a star in the S-star cluster would complete $\lsim 3\times 10^5$ orbits. Therefore, the cumulative heating (the left-hand side of equation~\ref{eq:cond1}) would be $\lsim 0.3\%$.

It is necessary to modify the above procedure for $40 M_{\odot}$ stars at the end of their lifetime ($\sim 4.5$ Myr). The right hand side of equation~\eqref{eq:rpAcc} is $2 \tilde{\omega}(-30)^{-2/3}=2.3$. Thus, the tidal coupling constant would become unreliable beyond $r_p/r_t=2.3$.
Secondly, there are numerical issues with high-order g-modes (i.e. the mode spectrum contains many gaps). Turning on the Cowling approximation (which neglects the perturbation to the stellar potential; \citealt{cowling1941}) ameliorates this problem, and allows us to include modes up to $\alpha=-90$. However, with the Cowling approximation equation~\eqref{eq:overlap} cannot be used. Instead, we use the original formulation of the overlap integral from \citet{press&teukolsky1977} for modes with $\alpha >-3$, and 
\begin{equation}
    Q_{\alpha,\ell}=\frac{\tilde{\omega}^2}{I_\alpha^{1/2}}\int_0^1 dx \tilde{\rho} x^{\ell+2} \left[\frac{\xi_{r, \alpha}}{\tilde{g}} +\frac{\xi_{h,\alpha}}{x^{\ell+1}} \left(\frac{x^{\ell+2}}{\tilde g}\right)' \right],
\end{equation}
for $\alpha\leq -3$, where $\tilde{g}$ is the acceleration of gravity in units of $G m_\star/r_\star^2$ (\citealt{ivanov+2013}; this formula is more numerically stable than the one in \citealt{press&teukolsky1977} for highly oscillatory modes).

\clearpage
\footnotesize{
\bibliographystyle{mnras}
\bibliography{master}
}

\bsp	
\label{lastpage}
\end{document}